\documentclass[preprint,aps]{revtex4}
\usepackage{graphicx}
\usepackage{dcolumn}
\usepackage{bm}
\usepackage{color}

\begin{document}

\preprint{APS/123-QED}

\title{Evolution of kink-like fluctuations associated with ion pickup within reconnection outflows in the Earth's magnetotail}

\author{Z. V\"or\"os and M.P. Leubner}
\affiliation{Institute of Astro- and Particle Physics, University of Innsbruck, Innsbruck, Austria}

\author{A. Runov and V. Angelopoulos}
\affiliation{Institute of Geophysics and Planetary Physics, UCLA, Los Angeles, USA}

\author{W. Baumjohann}
\affiliation{Space Research Institute, Austrian Academy of Sciences, Graz, Austria}

\date{\today}

\begin{abstract}
Magnetic reconnection (MR) in the Earth's magnetotail is usually followed by a system-wide redistribution of explosively released kinetic and thermal energy. Recently, multi-spacecraft observations from the THEMIS mission were used to study
localized explosions associated with MR in the magnetotail so as to understand
subsequent Earthward propagation of MR outbursts during substorms.
Here we investigate
plasma and magnetic field fluctuations/structures associated with MR exhaust and ion-ion kink mode instability during a well-documented THEMIS MR event.
Generation, evolution and fading of kink-like oscillations is followed over a distance of $\sim$ 70,000~km from the reconnection site in the mid- magnetotail to the more dipolar region near the Earth. We have found that the kink oscillations driven by
different ion populations within the outflow region can be
at least  
25,000~km from the reconnection site.
\end{abstract}

\maketitle

Collisionless magnetic reconnection (MR) necessitates formation of a thin current sheet in the magnetotail \cite{naka06}.
The current sheet 
can develop instabilities, including the kink mode. Kink-like flapping fluctuations were extensively observed by the four Cluster spacecraft \cite{serg03, run05}. Accordingly, the internally driven kink-like waves propagate from the center to the flanks of the magnetotail with a typical period of a few tens of s. Kink motions can be induced, for example, by the magnetic double gradient mechanism \cite{erka09} or by ion bulk speed shears, even under bifurcated current sheet configurations \cite{sitno04}.
The instabilities can also exhibit 
a large growth rate driven by  
relative streaming between separate ion populations. Though 
full particle and hybrid simulations with open boundary conditions \cite{kari03} revealed many salient features and observable signatures
of the ion-ion kink mode 
driving and evolution of associated fluctuations  
are not fully understood. Recent full particle simulations and 
analysis  indicate that the ion temperature in the reconnection outflow is proportional to 
proton bulk speed and 
ion species mass  \cite{drake09}.  
These results allow us in this paper to distinguish the
separate ion populations essential for driving the ion-ion kink. Moreover, we 
analyze the evolution of  MR outflow and  
kink-associated fluctuations using unique observations from the THEMIS mission.
The THEMIS mission, 
with its 
fleet of five identical spacecraft, was designed to study 
substorms during which a system-wide reconfiguration of the magnetosphere occurs. THEMIS timing analysis
(comparison of particle and field MR signatures at the
spacecraft positions) 
shows that system-wide substorm activity is initiated locally and the likely triggering mechanism is MR in the mid-magnetotail 
 \cite{ange08}.

We analyzed the field and plasma fluctuations associated with MR  
from 4.8 
to 5.3~UT on 26 February 2008 (hereafter, decimal hours will be used).
Substorm-related timing analysis of this event, together with ground-based (geomagnetic observatory), auroral and magnetotail signatures were  
described in great detail in \cite{ange08}. We used magnetic data 
with 3s time resolution from the THEMIS FGM  
experiment \cite{auster08} and 3s resolution ion moments computed from both ESA (energy range: eV - 25~keV) \cite{mcfad08} and SST (energy range: 25~keV - 6~MeV) instruments.
The THA spacecraft  
(TH - THEMIS; A-E, specific spacecraft),
situated at $\sim$5.5~$R_E$  (Earth radii), 
did not observe significant bulk flows. 
Because the observations from THD and THE, which were separated by $\sim 0.8~R_E$, were similar,
data from THA and THE will not be used further. The probes THB, THC and THD were aligned along X (from -21 to -11~$R_E$, see 
positions indicated  in Figure~1) within 3.9 $<$ Y $<$ 4.6 $R_E$ and -3 $<$ Z $<$ -2 $R_E$ in the GSM system of coordinates. 
Three probes 
were lined up from the mid-magnetotail to  
near-Earth space.
Since the average Y-Z cross-section of MR-driven bulk flows is 1-3~$R_E$ \cite{naka04}, 
the same Earthward flows could be observed by 
the spacecraft.
Figure~1 shows the magnetic field components $B_X, B_Z$ and the bulk speed component $V_X$  
in three vertically aligned subplots for each spacecraft. The subplots in the third row  
indicate that THB saw  
MR-associated flow reversal (before and after 05.0~h~UT: tailward - Earthward flows, $-+V_X$), 
whereas the other spacecraft observed predominantly Earthward flows. The flows reached maximum values of $|V_X| > 500$~km/s. The flow reversal is interpreted as movement  
of the reconnection X line through THB's position. 
Because THC approached the plasma sheet from the lobe ($B_X \leq -20$~nT), it did not observe the flow onset.
The Earthward flows were associated with fluctuations 
in  the $B_X$ and $B_Z$ magnetic components (
top two rows in Figure~1). Although the 
largest amplitude fluctuations in $B_X$ occurred closer to the MR site (at THB and THC),  they 
remained significant at THD. 
Since $B_X$ was frequently changing sign, the observed fluctuations could 
at least partially be attributed to a
flapping current sheet. The increase in fluctuations and magnitude of $B_Z$ (second row in Figure~1) during Earthward flows 
can be attributed to flow-driven dipolarizations of the magnetic field 
that are more significant closer to the Earth (at THD, last subplot).
\begin{figure}
\includegraphics[width=140mm]{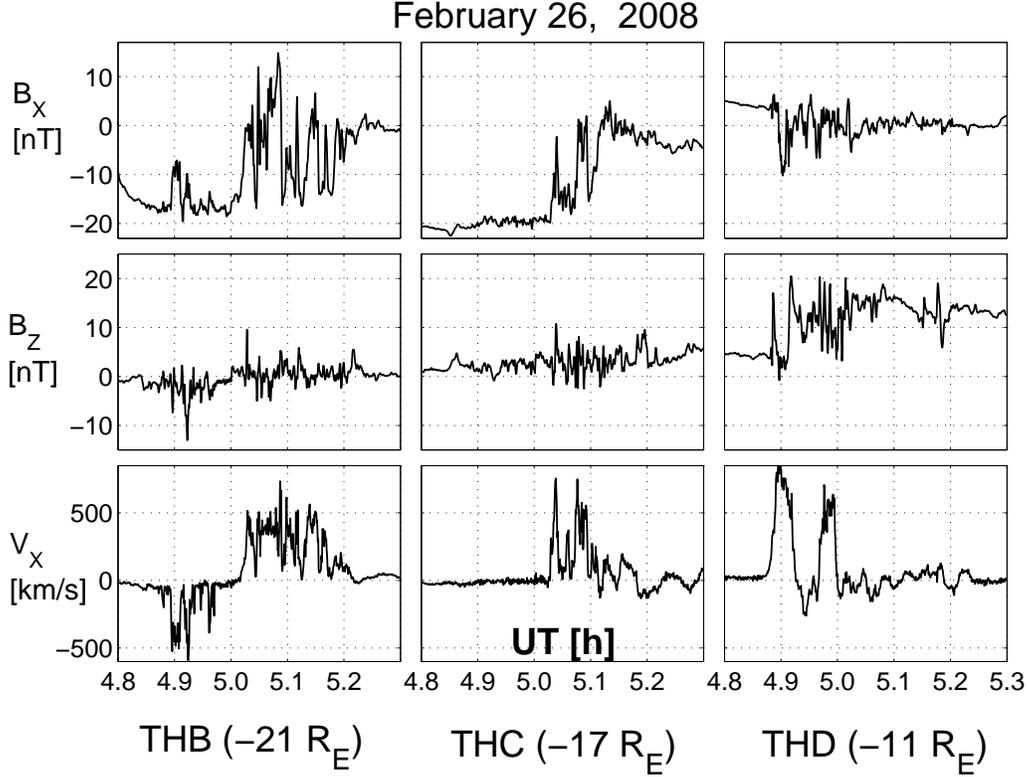}
\caption{\label{fig:location}
$B_X$, $B_Z$ magnetic and $V_X$ bulk speed GSM components observed by {\bf THEMIS:} 
THB (left), 
THC (middle), 
THD (right).}
\end{figure}

\begin{figure}
\includegraphics[width=140mm]{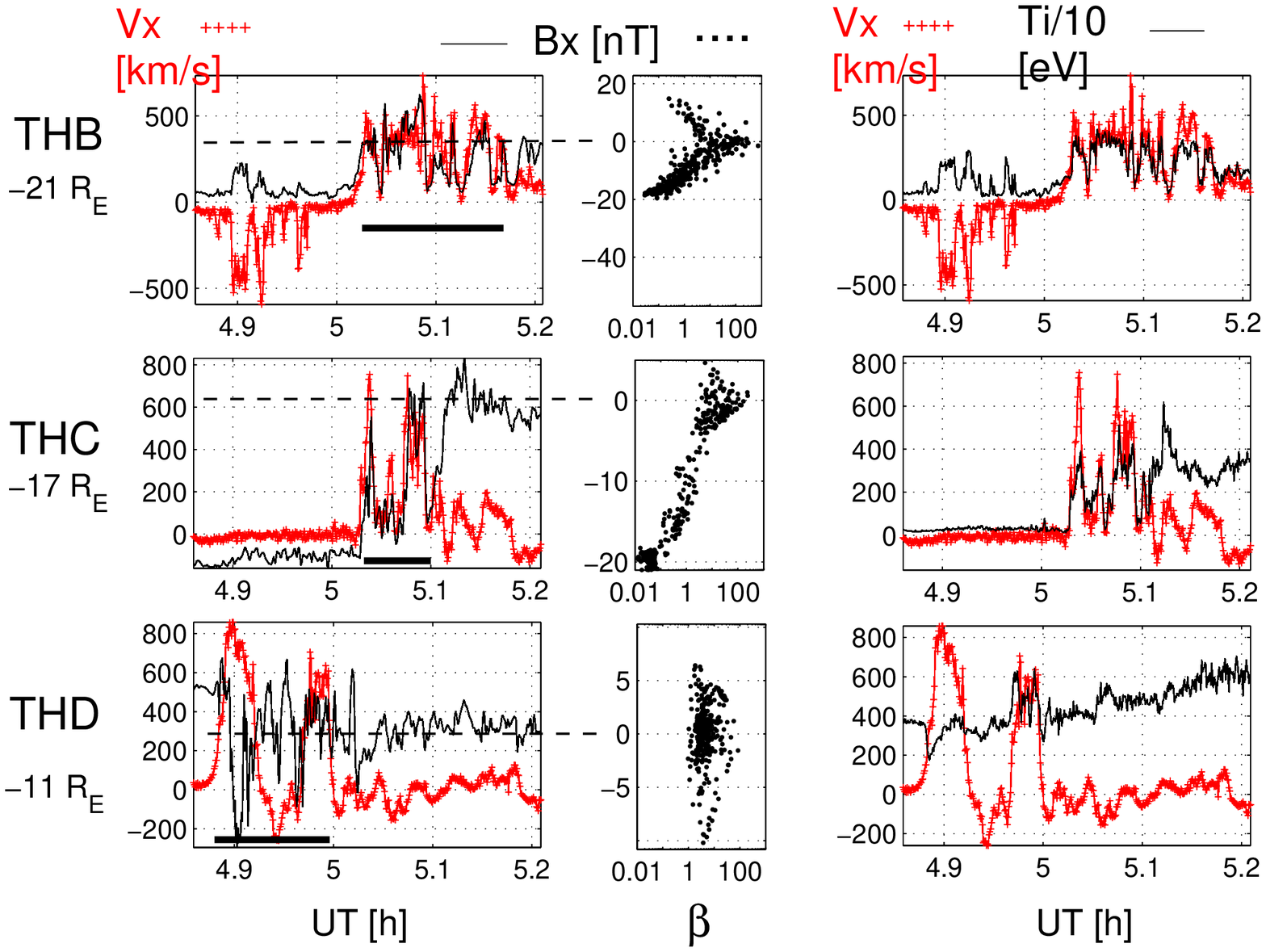}
\caption{\label{fig:location}
(Color). Correlations between $V_X$ and $B_X$ (left column);  
{\bf d}ependence of plasma $\beta$ on $B_X$ (middle column);  
{\bf c}orrelations between $V_X$ and $T_i/10 $; THB (top); THC (middle); THD (bottom).}
\end{figure}
Now we examine the orientation of current sheet oscillations associated with the 
$B_X$ sign change.
The current sheet orientation is estimated using minimum variance analysis (MVA), which is 
suitable for single spacecraft measurements. Assuming 
a 1-D boundary, the eigenvector ($\bf{ni}$) corresponding to the smallest eigenvalue ($\lambda_i$) computed from the covariance matrix of magnetic field is taken as the boundary normal \cite{sonne98}.
At THB and THC the flow-associated current sheet normal vectors pointed predominantly  
in the Y-Z directions ($n3y$ vs. $n3z$ pairs lay on a circle, not shown),  
$n3x$ was close to zero. This behavior is  characteristic  
of the Y-Z kink-like mode under tilted current sheet conditions \cite{serg06}. Nevertheless, this finding does not exclude the occurrence of a mixture of kink- and sausage-mode fluctuations.

MVA normals showed that at THD, the Y-Z kink signatures were largely lost, possibly because of different types of interactions during dipolarization and flow braking closer to the Earth.

Let us now investigate 
high-speed flow intervals
more closely. The first column in Figure~2 shows  
superimposed $V_X$ and $B_X$ fluctuations for each spacecraft. The
axis values
for $B_X$
are the same for the first and second columns
in the figure.
The second column shows $B_X$ as a function of plasma $\beta$. 
The horizontal dashed lines represent 
$B_X=0$ nT; 
the horizontal thick lines, 
high-speed flow intervals.  
The correlation coefficients between $B_X$ and $V_X$ fluctuations were 0.66 and 0.68 at THB and THC, respectively.
The correlations between current sheet flapping motions ($\pm B_X$) and bulk speed ($V_X$) fluctuations can be partially explained through a spatial effect:  
plasma $\beta$ increases towards the center of the current sheet and towards the central part of the plasma flows. In fact,
the condition $\beta > 2$ indicates that a spacecraft is merged to a high-speed bulk flow \cite{naka04}.
As the probe moves through the current sheet, it
can intersect different parts of the flow again and again. The correlation between $B_X$ and $T_i$ (not shown) emerges
for the same reason. The right column shows  
superimposed  $V_X$ and $Ti/10$ fluctuations for each spacecraft. The correlation coefficients are higher between the plasma parameters 0.91 and 0.86 
than between the magnetic field and plasma parameters 0.66 and 0.68 
for THB and THC, respectively. The fluctuations at THD  are uncorrelated, since 
the spacecraft are in the center of the plasma sheet in the dipolarization region, where plasma $\beta$ is changing less. Along with MVA analysis results, the figure 
shows that magnetic and plasma fluctuations evidencing of kink oscillations span a distance of at least  $\sim 4 R_E$ (from THB to THC), 
but do not reach the position of THD. A similar large-scale extension of kink-like oscillations in the magnetotail was observed only once during a conjunction of Cluster and Double Star satellites. \cite{zhang05}.
\begin{figure}
\includegraphics[width=140mm]{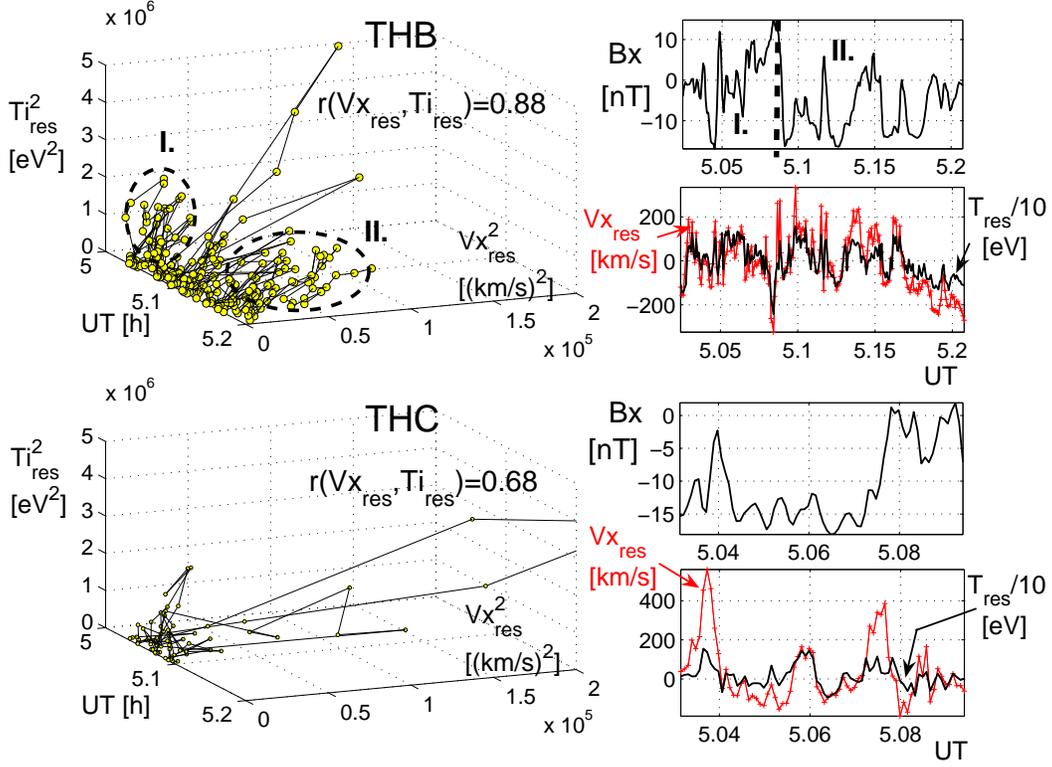}
\caption{\label{fig:loation}
(Color). Comparison of the time evolution of plasma residuals and the flapping magnetic field at THB (top) and at THC (bottom).}
\end{figure}

An important question is what drives 
speed and magnetic kink-like fluctuations during the high-speed MR outflows at THB and THC. Global MHD simulations show that directional changes in 
solar wind speed (mainly in $V_Z$) can induce neutral sheet flapping 
in the near and middle magnetotail 
with a delay of
10 to 
15 minutes \cite{serg08}
We have checked time-shifted (to the distance of the Earth) ACE solar wind data between 04.66 and 
05.66~UT and found the following mean and standard deviations: $V_X=-385\pm2$ km/s, $V_Y=-10\pm3$ km/s, $V_Z=32\pm3$ km/s, $B_X=0.3\pm1$ nT, $B_Y=2.2\pm1$ nT and $B_Z=1.2\pm0.3$ nT. Since the changes are 
insignificant, flapping current sheet motions are unlikely to be driven by the solar wind. Moreover, the current sheet is quiet  
between tailward and Earthward flows ( 
top left subplot in Figure 2). Intense current sheet oscillations and fast flows arose 
simultaneously. If the solar wind were 
driving the current sheet, strong $B_X$ oscillations would be present  
between the tailward and 
Earthward flow regions, as well. Current sheet oscillations  
due to finite $B_Y$ in the neutral sheet or a number of instabilities \cite{birn07} can occur during the entire  
reconnection interval \cite{lait07}.
We are not aware of any experimental work  that  
clearly identifies a single 
mechanism driving  
current sheet oscillations in space. We 
show here that 
different ion populations exist near MR (at THB), and 
strong interactions between the magnetic field and plasma outflow are present near and at least 4~$R_E$ away from the MR site. These interactions  could 
be responsible for current sheet kink oscillations during the fast flows. Since the correlations between 
$B_X$, $V_X$ and $T_i$ during fast flows are lost at THD (see the bottom row of subplots in Figure 2), we consider only the fluctuations at THB and THC.  
To remove the spatial effects due to multiple crossings of the current sheet and plasma flows (for which the changing plasma $\beta$ serves as a proxy measure), we evaluated the linear fit between $B_X$, $V_X$ and $B_X$, $T_i$ during fast flows in a least-squares sense, respectively. We suppose, the residuals of the fit correspond to the fluctuations of plasma parameters not influenced by the flapping current sheet. The physical significance of residuals is also supported by
the higher correlation between plasma parameters 
than between plasma and $B_X$ fluctuations; 
that is, 
plasma parameter fluctuations
are not determined exclusively by
flapping. Figure~3 shows the
relationship 
between the residuals from the linear fit ($Vx_{res}, Ti_{res}$) and $B_X$ at THB (top panels) and THC (bottom panels).  On the right,  $B_X$,  $Vx_{res}$ and $Ti_{res}/10$ time series are depicted. On the left, the 3D panel shows the 
the squares of the speed and temperature residuals as functions of time.

\begin{figure}
\includegraphics[width=140mm]{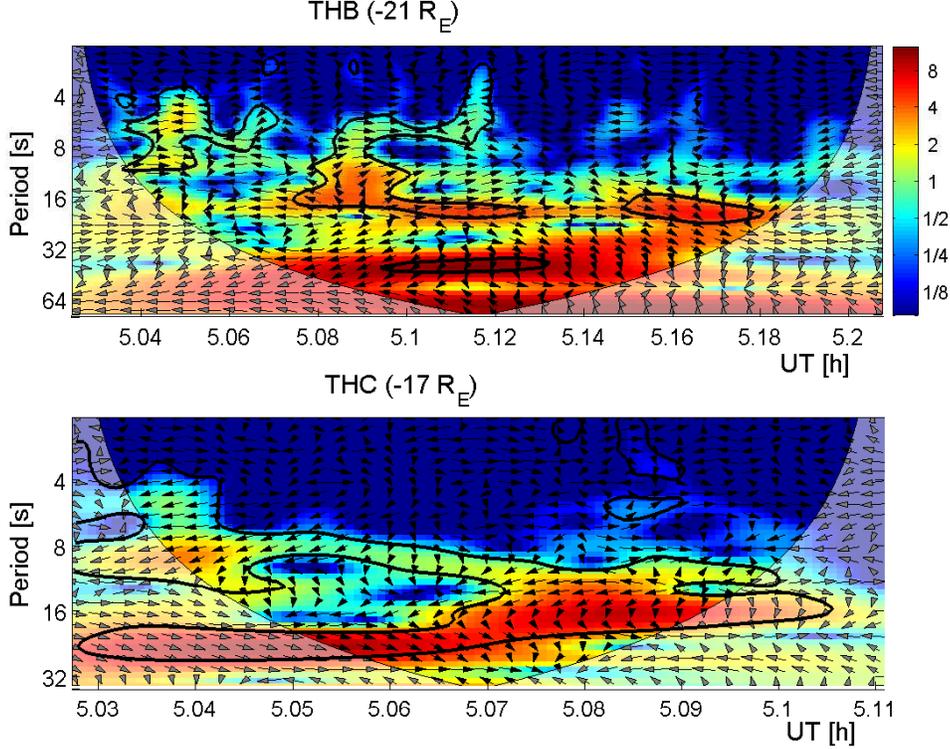}
\caption{\label{fig:tauq}
(Color).Cross-wavelet spectra between $Vx_{res}$ and $B_X$ at THB (top) and THC (bottom);
colors indicate cross-wavelet power; thick contours represent 5\% significant level against red noise; the arrows correspond to the cross-wavelet phase: with in-phase (anti-phase) pointing right (left), and $Vx_{res}$ ($B_X$) leading $B_X$ ($Vx_{res}$) by 90$^o$ pointing down (up). The lighter shade indicates regions where edge effects might distort the spectra.}
\end{figure}

There are several position-dependent items of interest.
Close to MR (at THB) the residuals (top-left in Figure~3) show two well-correlated plasma populations,  
I and II.
Population~I,  characterized by its low speed and high temperature, appeared before $\sim$ 05.09 UT; Population~II, characterized by its
high speed and low temperature, appeared after  $\sim$ 05.09 UT.
Recent full-particle simulations indicate that ions crossing the MR outflow boundary can behave like non-adiabatic pickup particles, gaining an effective thermal speed depending on the bulk speed of
the reconnection exhaust.
The total ion temperature depends on the square of the outflow speed and the mass of ion species
\cite{drake09}.
Cluster analysis has already revealed that ion motion is usually non-adiabatic in the flapping current sheets \cite{run06}.
This particle pickup mechanism represents a
possible explanation of the observed 
speed-temperature correlation. 
The combination of Y-Z like kink motions (see above) and the deduced change in ion populations indicate
the occurrence of
ion-ion kink instability. Moreover, we show that the dynamics of associated current sheet flapping changes when the plasma parameters change from population I to population II. 
The correlation coefficient between the residuals is $r(Vx_{res}, Ti_{res})=0.87$, and the effect of flapping is largely removed, since $r(Vx_{res}, B_X)\sim r(Ti_{res},B_X)\sim0.02$.  Nevertheless, the cross-wavelet power
between  
$Vx_{res}$ (or $Ti_{res}$, not shown) and $B_X$ 
is significant
(top subplot in Figure~4). The cross-wavelet plot associated with population I  plasma (before $\sim$ 05.09 UT) displays 
significant power over a range of periods ($\sim$ 4-20 s), while longer periods ($\sim$ 15-60 s) are associated with the population II plasma (after $\sim$ 05.09 UT). An inspection of $B_X$ data (right in Figure 3) also indicates that the character of $B_X$ fluctuations 
changes when the border (a vertical line at 05.09 UT) between plasma populations I and II is crossed. Shorter-period magnetic fluctuations ($\sim$ 10-70 s) are replaced by longer-period ($\sim$ 70-120 s) ones. The longer periods represent typical periodicities found in 3D full particle and hybrid simulations during the late stage of ion-ion kink instability \cite{kari03} and a characteristic period observed by Cluster during current sheet kink oscillations  
\cite{runo03}. Simulation results also show that short period ion-ion kink oscillations occur first, evolving later into longer period modes \cite{kari03}. The transition from shorter- to longer-period fluctuations is visible in both $B_X$ (Figure3) and cross-wavelet spectra (Figure 4) at THB. At THB and THC positions, the total magnetic field (not shown) strongly fluctuates between 2-17 nT. The corresponding proton gyroperiod varies between 4-20 s, overlapping the range of observed shorter-period fluctuations.

At the position of THC the correlation between the plasma parameter residuals and magnetic field ($r(Vx_{res}, B_X)\sim r(Ti_{res},B_X)\sim0.04$) is negligible. 
At a distance of 4 $R_E$ Earthward from MR, the interconnection between residuals is weaker ($r(Vx_{res}, Ti_{res})=0.68$). Though THC is 
in the flow for less time, 
no differing plasma populations are visible, and the cross-wavelet $Vx_{res}$, $B_X$ plot shows common power at discrete periods of $\sim$8 and 25 s. $B_X$ shows the largest power over periods of 100-200 s (not shown). The decreasing correlations and increasing periods at THC signal that the ion-ion kink reached its saturation or even fading phase.
The wavelet phase, which would reveal the in-phase or out-of-phase features \cite{agrin04} associated with the quantities in the cross-wavelet plane, exhibits knotty,  period-dependent patterns at THB (the direction of arrows in Figure 4). The pattern is less complex at THC. Nevertheless, it seems to be impossible to identify a clear wavelet phase between plasma and magnetic field fluctuations. We can only say 
that the cross-wavelet fluctuations exhibit compound multi-scale interactions and the cross-power of these interactions is changing with time and position.

In summary, we have identified reconnection outflow-associated plasma and magnetic field correlations in the presence of ion-ion kink mode oscillations of the current sheet. The evolution and fading of the kink mode with distance 
are connected with decreasing speed-temperature correlations, therefore with decreasing efficiency of pickup ion feeding.
These findings suggest that the simultaneous occurrence of multi-scale interactions
between the flapping magnetic field and the plasma flow can partially be responsible for
the gradual disappearance of kink mode oscillations. Further analysis is needed, however,
to establish causal connections between driving and dissipation mechanisms for the observed
correlated fluctuations.
\begin{acknowledgments}
We 
thank V. Sergeev for 
helpful discussions and J. Hohl for spelling corrections.
The work of Z.V. and M.P.L.  was supported  by the  Austrian
Wissenschaftsfonds (FWF) under grant ~P20131-N16. Cross-wavelet software was provided by
A. Grinsted. We acknowledge support from
NASA NAS5-02099.

\end{acknowledgments}


\begin{thebibliography}{20}
\expandafter\ifx\csname natexlab\endcsname\relax\def\natexlab#1{#1}\fi
\expandafter\ifx\csname bibnamefont\endcsname\relax
  \def\bibnamefont#1{#1}\fi
\expandafter\ifx\csname bibfnamefont\endcsname\relax
  \def\bibfnamefont#1{#1}\fi
\expandafter\ifx\csname citenamefont\endcsname\relax
  \def\citenamefont#1{#1}\fi
\expandafter\ifx\csname url\endcsname\relax
  \def\url#1{\texttt{#1}}\fi
\expandafter\ifx\csname urlprefix\endcsname\relax\def\urlprefix{URL }\fi
\providecommand{\bibinfo}[2]{#2}
\providecommand{\eprint}[2][]{\url{#2}}

\bibitem[{\citenamefont{Nakamura et~al.}(2006)\citenamefont{Nakamura,
  Baumjohann, Asano, Runov, Balogh, Owen, Fazakerley, Fujimoto, Klecker, and
  R\`eme}}]{naka06}
\bibinfo{author}{\bibfnamefont{R.}~\bibnamefont{Nakamura}},
  \bibinfo{author}{\bibfnamefont{W.}~\bibnamefont{Baumjohann}},
  \bibinfo{author}{\bibfnamefont{Y.}~\bibnamefont{Asano}},
  \bibinfo{author}{\bibfnamefont{A.}~\bibnamefont{Runov}},
  \bibinfo{author}{\bibfnamefont{A.}~\bibnamefont{Balogh}},
  \bibinfo{author}{\bibfnamefont{C.}~\bibnamefont{Owen}},
  \bibinfo{author}{\bibfnamefont{A.}~\bibnamefont{Fazakerley}},
  \bibinfo{author}{\bibfnamefont{M.}~\bibnamefont{Fujimoto}},
  \bibinfo{author}{\bibfnamefont{B.}~\bibnamefont{Klecker}}, \bibnamefont{and}
  \bibinfo{author}{\bibfnamefont{H.}~\bibnamefont{R\`eme}},
  \bibinfo{journal}{J.\ Geophys.\ Res.} \textbf{\bibinfo{volume}{111}},
  \bibinfo{pages}{A11206} (\bibinfo{year}{2006}).

\bibitem[{\citenamefont{Sergeev et~al.}(2003)\citenamefont{Sergeev, Runov,
  Baumjohann, Nakamura, Zhang, Volwerk, Balohg, Reme, Sauvaud, Andre
  et~al.}}]{serg03}
\bibinfo{author}{\bibfnamefont{V.}~\bibnamefont{Sergeev}},
  \bibinfo{author}{\bibfnamefont{A.}~\bibnamefont{Runov}},
  \bibinfo{author}{\bibfnamefont{W.}~\bibnamefont{Baumjohann}},
  \bibinfo{author}{\bibfnamefont{R.}~\bibnamefont{Nakamura}},
  \bibinfo{author}{\bibfnamefont{T.}~\bibnamefont{Zhang}},
  \bibinfo{author}{\bibfnamefont{M.}~\bibnamefont{Volwerk}},
  \bibinfo{author}{\bibfnamefont{A.}~\bibnamefont{Balohg}},
  \bibinfo{author}{\bibfnamefont{H.}~\bibnamefont{Reme}},
  \bibinfo{author}{\bibfnamefont{J.}~\bibnamefont{Sauvaud}},
  \bibinfo{author}{\bibfnamefont{M.}~\bibnamefont{Andre}},
  \bibnamefont{et~al.}, \bibinfo{journal}{Geophys.\ Res.\ Lett.}
  \textbf{\bibinfo{volume}{30}}, \bibinfo{pages}{1327} (\bibinfo{year}{2003}).

\bibitem[{\citenamefont{Runov et~al.}(2005)\citenamefont{Runov, Sergeev,
  Baumjohann, Nakamura, Apatenkov, Asano, Volwerk, V\"{o}r\"{o}s, Zhang,
  Petrukovich et~al.}}]{run05}
\bibinfo{author}{\bibfnamefont{A.}~\bibnamefont{Runov}},
  \bibinfo{author}{\bibfnamefont{V.}~\bibnamefont{Sergeev}},
  \bibinfo{author}{\bibfnamefont{W.}~\bibnamefont{Baumjohann}},
  \bibinfo{author}{\bibfnamefont{R.}~\bibnamefont{Nakamura}},
  \bibinfo{author}{\bibfnamefont{S.}~\bibnamefont{Apatenkov}},
  \bibinfo{author}{\bibfnamefont{Y.}~\bibnamefont{Asano}},
  \bibinfo{author}{\bibfnamefont{M.}~\bibnamefont{Volwerk}},
  \bibinfo{author}{\bibfnamefont{Z.}~\bibnamefont{V\"{o}r\"{o}s}},
  \bibinfo{author}{\bibfnamefont{T.}~\bibnamefont{Zhang}},
  \bibinfo{author}{\bibfnamefont{A.}~\bibnamefont{Petrukovich}},
  \bibnamefont{et~al.}, \bibinfo{journal}{Ann.\ Geophys.}
  \textbf{\bibinfo{volume}{23}}, \bibinfo{pages}{1391} (\bibinfo{year}{2005}).

\bibitem[{\citenamefont{Erkaev et~al.}(2009)\citenamefont{Erkaev, Semenov,
  Kubyshkin, Kubyshkina, and Biernat}}]{erka09}
\bibinfo{author}{\bibfnamefont{N.}~\bibnamefont{Erkaev}},
  \bibinfo{author}{\bibfnamefont{V.}~\bibnamefont{Semenov}},
  \bibinfo{author}{\bibfnamefont{I.}~\bibnamefont{Kubyshkin}},
  \bibinfo{author}{\bibfnamefont{M.}~\bibnamefont{Kubyshkina}},
  \bibnamefont{and} \bibinfo{author}{\bibfnamefont{H.}~\bibnamefont{Biernat}},
  \bibinfo{journal}{Ann.\ Geophys.} \textbf{\bibinfo{volume}{27}},
  \bibinfo{pages}{417} (\bibinfo{year}{2009}).

\bibitem[{\citenamefont{Sitnov et~al.}(2004)\citenamefont{Sitnov, Swisdak,
  Drake, Guzdar, and Rogers}}]{sitno04}
\bibinfo{author}{\bibfnamefont{M.}~\bibnamefont{Sitnov}},
  \bibinfo{author}{\bibfnamefont{M.}~\bibnamefont{Swisdak}},
  \bibinfo{author}{\bibfnamefont{J.}~\bibnamefont{Drake}},
  \bibinfo{author}{\bibfnamefont{P.}~\bibnamefont{Guzdar}}, \bibnamefont{and}
  \bibinfo{author}{\bibfnamefont{B.}~\bibnamefont{Rogers}},
  \bibinfo{journal}{Geophys.\ Res.\ Lett.} \textbf{\bibinfo{volume}{31}},
  \bibinfo{pages}{L09805} (\bibinfo{year}{2004}).

\bibitem[{\citenamefont{Karimabadi et~al.}(2003)\citenamefont{Karimabadi,
  Pritchett, Daughton, and Krauss-Varban}}]{kari03}
\bibinfo{author}{\bibfnamefont{H.}~\bibnamefont{Karimabadi}},
  \bibinfo{author}{\bibfnamefont{P.}~\bibnamefont{Pritchett}},
  \bibinfo{author}{\bibfnamefont{W.}~\bibnamefont{Daughton}}, \bibnamefont{and}
  \bibinfo{author}{\bibfnamefont{D.}~\bibnamefont{Krauss-Varban}},
  \bibinfo{journal}{J.\ Geophys.\ Res.} \textbf{\bibinfo{volume}{108(A11)}},
  \bibinfo{pages}{1401} (\bibinfo{year}{2003}).

\bibitem[{\citenamefont{Drake et~al.}(2009)\citenamefont{Drake, Swisdak, Phan,
  Cassak, Shay, Lepri, Lin, Quataert, and Zurbuchen}}]{drake09}
\bibinfo{author}{\bibfnamefont{J.}~\bibnamefont{Drake}},
  \bibinfo{author}{\bibfnamefont{M.}~\bibnamefont{Swisdak}},
  \bibinfo{author}{\bibfnamefont{T.}~\bibnamefont{Phan}},
  \bibinfo{author}{\bibfnamefont{P.}~\bibnamefont{Cassak}},
  \bibinfo{author}{\bibfnamefont{M.}~\bibnamefont{Shay}},
  \bibinfo{author}{\bibfnamefont{S.}~\bibnamefont{Lepri}},
  \bibinfo{author}{\bibfnamefont{R.}~\bibnamefont{Lin}},
  \bibinfo{author}{\bibfnamefont{E.}~\bibnamefont{Quataert}}, \bibnamefont{and}
  \bibinfo{author}{\bibfnamefont{T.}~\bibnamefont{Zurbuchen}},
  \bibinfo{journal}{J.\ Geophys.\ Res.} \textbf{\bibinfo{volume}{114}},
  \bibinfo{pages}{A05111} (\bibinfo{year}{2009}).

\bibitem[{\citenamefont{Angelopoulos et~al.}(2008)\citenamefont{Angelopoulos,
  McFadden, Larson, Carlson, Mende, Frey, Phan, Sibeck, Glassmeier, Auster
  et~al.}}]{ange08}
\bibinfo{author}{\bibfnamefont{V.}~\bibnamefont{Angelopoulos}},
  \bibinfo{author}{\bibfnamefont{J.}~\bibnamefont{McFadden}},
  \bibinfo{author}{\bibfnamefont{D.}~\bibnamefont{Larson}},
  \bibinfo{author}{\bibfnamefont{C.}~\bibnamefont{Carlson}},
  \bibinfo{author}{\bibfnamefont{S.}~\bibnamefont{Mende}},
  \bibinfo{author}{\bibfnamefont{H.}~\bibnamefont{Frey}},
  \bibinfo{author}{\bibfnamefont{T.}~\bibnamefont{Phan}},
  \bibinfo{author}{\bibfnamefont{D.}~\bibnamefont{Sibeck}},
  \bibinfo{author}{\bibfnamefont{K.-H.} \bibnamefont{Glassmeier}},
  \bibinfo{author}{\bibfnamefont{U.}~\bibnamefont{Auster}},
  \bibnamefont{et~al.}, \bibinfo{journal}{Science}
  \textbf{\bibinfo{volume}{321}}, \bibinfo{pages}{931} (\bibinfo{year}{2008}).

\bibitem[{\citenamefont{Auster et~al.}(2008)\citenamefont{Auster, Glassmeier,
  Magnes, Aydogar, Baumjohann, Constantinescu, Fischer, Fornacon, Georgescu,
  Harvey et~al.}}]{auster08}
\bibinfo{author}{\bibfnamefont{H.}~\bibnamefont{Auster}},
  \bibinfo{author}{\bibfnamefont{K.}~\bibnamefont{Glassmeier}},
  \bibinfo{author}{\bibfnamefont{W.}~\bibnamefont{Magnes}},
  \bibinfo{author}{\bibfnamefont{O.}~\bibnamefont{Aydogar}},
  \bibinfo{author}{\bibfnamefont{W.}~\bibnamefont{Baumjohann}},
  \bibinfo{author}{\bibfnamefont{D.}~\bibnamefont{Constantinescu}},
  \bibinfo{author}{\bibfnamefont{D.}~\bibnamefont{Fischer}},
  \bibinfo{author}{\bibfnamefont{K.}~\bibnamefont{Fornacon}},
  \bibinfo{author}{\bibfnamefont{E.}~\bibnamefont{Georgescu}},
  \bibinfo{author}{\bibfnamefont{P.}~\bibnamefont{Harvey}},
  \bibnamefont{et~al.}, \bibinfo{journal}{Space Sci.\ Rev.}
  \textbf{\bibinfo{volume}{141}}, \bibinfo{pages}{235} (\bibinfo{year}{2008}).

\bibitem[{\citenamefont{McFadden et~al.}(2008)\citenamefont{McFadden, Carlson,
  Larson, Ludlam, Abiad, Elliott, Turin, Marckwordt, and
  Angelopoulos}}]{mcfad08}
\bibinfo{author}{\bibfnamefont{J.}~\bibnamefont{McFadden}},
  \bibinfo{author}{\bibfnamefont{C.}~\bibnamefont{Carlson}},
  \bibinfo{author}{\bibfnamefont{D.}~\bibnamefont{Larson}},
  \bibinfo{author}{\bibfnamefont{M.}~\bibnamefont{Ludlam}},
  \bibinfo{author}{\bibfnamefont{R.}~\bibnamefont{Abiad}},
  \bibinfo{author}{\bibfnamefont{B.}~\bibnamefont{Elliott}},
  \bibinfo{author}{\bibfnamefont{P.}~\bibnamefont{Turin}},
  \bibinfo{author}{\bibfnamefont{M.}~\bibnamefont{Marckwordt}},
  \bibnamefont{and}
  \bibinfo{author}{\bibfnamefont{V.}~\bibnamefont{Angelopoulos}},
  \bibinfo{journal}{Space Sci.\ Rev.} \textbf{\bibinfo{volume}{141}},
  \bibinfo{pages}{277} (\bibinfo{year}{2008}).

\bibitem[{\citenamefont{Nakamura et~al.}(2004)\citenamefont{Nakamura,
  Baumjohann, Mouikis, Kistler, Runov, Volwerk, Asano, V\"or\"os, Zhang,
  Klecker et~al.}}]{naka04}
\bibinfo{author}{\bibfnamefont{R.}~\bibnamefont{Nakamura}},
  \bibinfo{author}{\bibfnamefont{W.}~\bibnamefont{Baumjohann}},
  \bibinfo{author}{\bibfnamefont{C.}~\bibnamefont{Mouikis}},
  \bibinfo{author}{\bibfnamefont{L.}~\bibnamefont{Kistler}},
  \bibinfo{author}{\bibfnamefont{A.}~\bibnamefont{Runov}},
  \bibinfo{author}{\bibfnamefont{M.}~\bibnamefont{Volwerk}},
  \bibinfo{author}{\bibfnamefont{Y.}~\bibnamefont{Asano}},
  \bibinfo{author}{\bibfnamefont{Z.}~\bibnamefont{V\"or\"os}},
  \bibinfo{author}{\bibfnamefont{T.}~\bibnamefont{Zhang}},
  \bibinfo{author}{\bibfnamefont{B.}~\bibnamefont{Klecker}},
  \bibnamefont{et~al.}, \bibinfo{journal}{Geophys.\ Res.\ Lett.}
  \textbf{\bibinfo{volume}{31}}, \bibinfo{pages}{L09804}
  (\bibinfo{year}{2004}).

\bibitem[{\citenamefont{Sonnerup and Scheible}(2007)}]{sonne98}
\bibinfo{author}{\bibfnamefont{B.}~\bibnamefont{Sonnerup}} \bibnamefont{and}
  \bibinfo{author}{\bibfnamefont{M.}~\bibnamefont{Scheible}}, in
  \emph{\bibinfo{booktitle}{Analysis methods for multi-spacecraft data}}
  (\bibinfo{publisher}{ISSI/ESA}, \bibinfo{year}{2007}), p.
  \bibinfo{pages}{185}.

\bibitem[{\citenamefont{Sergeev et~al.}(2006)\citenamefont{Sergeev, Sormakov,
  Apatenkov, Baumjohann, Nakamura, Runov, Mukai, and Nagai}}]{serg06}
\bibinfo{author}{\bibfnamefont{V.}~\bibnamefont{Sergeev}},
  \bibinfo{author}{\bibfnamefont{D.}~\bibnamefont{Sormakov}},
  \bibinfo{author}{\bibfnamefont{S.}~\bibnamefont{Apatenkov}},
  \bibinfo{author}{\bibfnamefont{W.}~\bibnamefont{Baumjohann}},
  \bibinfo{author}{\bibfnamefont{R.}~\bibnamefont{Nakamura}},
  \bibinfo{author}{\bibfnamefont{A.}~\bibnamefont{Runov}},
  \bibinfo{author}{\bibfnamefont{T.}~\bibnamefont{Mukai}}, \bibnamefont{and}
  \bibinfo{author}{\bibfnamefont{T.}~\bibnamefont{Nagai}},
  \bibinfo{journal}{Ann.\ Geophys.} \textbf{\bibinfo{volume}{24}},
  \bibinfo{pages}{2015} (\bibinfo{year}{2006}).

\bibitem[{\citenamefont{Zhang et~al.}(2005)\citenamefont{Zhang, Nakamura,
  Volwerk, Runov, Baumjohann, Eichelberger, Carr, Balogh, Sergeev, Shi
  et~al.}}]{zhang05}
\bibinfo{author}{\bibfnamefont{T.}~\bibnamefont{Zhang}},
  \bibinfo{author}{\bibfnamefont{R.}~\bibnamefont{Nakamura}},
  \bibinfo{author}{\bibfnamefont{M.}~\bibnamefont{Volwerk}},
  \bibinfo{author}{\bibfnamefont{A.}~\bibnamefont{Runov}},
  \bibinfo{author}{\bibfnamefont{W.}~\bibnamefont{Baumjohann}},
  \bibinfo{author}{\bibfnamefont{H.}~\bibnamefont{Eichelberger}},
  \bibinfo{author}{\bibfnamefont{C.}~\bibnamefont{Carr}},
  \bibinfo{author}{\bibfnamefont{A.}~\bibnamefont{Balogh}},
  \bibinfo{author}{\bibfnamefont{V.}~\bibnamefont{Sergeev}},
  \bibinfo{author}{\bibfnamefont{J.~K.} \bibnamefont{Shi}},
  \bibnamefont{et~al.}, \bibinfo{journal}{Ann.\ Geophys.}
  \textbf{\bibinfo{volume}{23}}, \bibinfo{pages}{2909} (\bibinfo{year}{2005}).

\bibitem[{\citenamefont{Sergeev et~al.}(2008)\citenamefont{Sergeev, Tsyganenko,
  and Angelopoulos}}]{serg08}
\bibinfo{author}{\bibfnamefont{V.}~\bibnamefont{Sergeev}},
  \bibinfo{author}{\bibfnamefont{N.}~\bibnamefont{Tsyganenko}},
  \bibnamefont{and}
  \bibinfo{author}{\bibfnamefont{V.}~\bibnamefont{Angelopoulos}},
  \bibinfo{journal}{Ann.\ Geophys.} \textbf{\bibinfo{volume}{26}},
  \bibinfo{pages}{2395} (\bibinfo{year}{2008}).

\bibitem[{\citenamefont{Birn and Priest}(2007)}]{birn07}
\bibinfo{editor}{\bibfnamefont{J.}~\bibnamefont{Birn}} \bibnamefont{and}
  \bibinfo{editor}{\bibfnamefont{E.}~\bibnamefont{Priest}}, eds.,
  \emph{\bibinfo{title}{Reconnection of magnetic fields}}
  (\bibinfo{publisher}{Cambridge Univ. Press}, \bibinfo{year}{2007}).

\bibitem[{\citenamefont{Laitinen et~al.}(2007)\citenamefont{Laitinen, Nakamura,
  Runov, R\`eme, and Lucek}}]{lait07}
\bibinfo{author}{\bibfnamefont{T.}~\bibnamefont{Laitinen}},
  \bibinfo{author}{\bibfnamefont{R.}~\bibnamefont{Nakamura}},
  \bibinfo{author}{\bibfnamefont{A.}~\bibnamefont{Runov}},
  \bibinfo{author}{\bibfnamefont{H.}~\bibnamefont{R\`eme}}, \bibnamefont{and}
  \bibinfo{author}{\bibfnamefont{E.}~\bibnamefont{Lucek}},
  \bibinfo{journal}{Ann.\ Geophys.} \textbf{\bibinfo{volume}{25}},
  \bibinfo{pages}{1025} (\bibinfo{year}{2007}).

\bibitem[{\citenamefont{Runov et~al.}(2006)\citenamefont{Runov, Sergeev,
  Nakamura, Baumjohann, Apatenkov, Asano, Takada, Volwerk, V\"{o}r\"{o}s, Zhang
  et~al.}}]{run06}
\bibinfo{author}{\bibfnamefont{A.}~\bibnamefont{Runov}},
  \bibinfo{author}{\bibfnamefont{V.}~\bibnamefont{Sergeev}},
  \bibinfo{author}{\bibfnamefont{R.}~\bibnamefont{Nakamura}},
  \bibinfo{author}{\bibfnamefont{W.}~\bibnamefont{Baumjohann}},
  \bibinfo{author}{\bibfnamefont{S.}~\bibnamefont{Apatenkov}},
  \bibinfo{author}{\bibfnamefont{Y.}~\bibnamefont{Asano}},
  \bibinfo{author}{\bibfnamefont{T.}~\bibnamefont{Takada}},
  \bibinfo{author}{\bibfnamefont{M.}~\bibnamefont{Volwerk}},
  \bibinfo{author}{\bibfnamefont{Z.}~\bibnamefont{V\"{o}r\"{o}s}},
  \bibinfo{author}{\bibfnamefont{T.}~\bibnamefont{Zhang}},
  \bibnamefont{et~al.}, \bibinfo{journal}{Ann.\ Geophys.}
  \textbf{\bibinfo{volume}{24}}, \bibinfo{pages}{247} (\bibinfo{year}{2006}).

\bibitem[{\citenamefont{Runov et~al.}(2003)\citenamefont{Runov, Nakamura,
  Baumjohann, Zhang, Volwerk, Eichelberger, and Balogh}}]{runo03}
\bibinfo{author}{\bibfnamefont{A.}~\bibnamefont{Runov}},
  \bibinfo{author}{\bibfnamefont{R.}~\bibnamefont{Nakamura}},
  \bibinfo{author}{\bibfnamefont{W.}~\bibnamefont{Baumjohann}},
  \bibinfo{author}{\bibfnamefont{T.}~\bibnamefont{Zhang}},
  \bibinfo{author}{\bibfnamefont{M.}~\bibnamefont{Volwerk}},
  \bibinfo{author}{\bibfnamefont{H.}~\bibnamefont{Eichelberger}},
  \bibnamefont{and} \bibinfo{author}{\bibfnamefont{A.}~\bibnamefont{Balogh}},
  \bibinfo{journal}{Geophys.\ Res.\ Lett.} \textbf{\bibinfo{volume}{30(2)}},
  \bibinfo{pages}{1036} (\bibinfo{year}{2003}).

\bibitem[{\citenamefont{Grinsted et~al.}(2004)\citenamefont{Grinsted, Moore,
  and Jevrejeva}}]{agrin04}
\bibinfo{author}{\bibfnamefont{A.}~\bibnamefont{Grinsted}},
  \bibinfo{author}{\bibfnamefont{J.}~\bibnamefont{Moore}}, \bibnamefont{and}
  \bibinfo{author}{\bibfnamefont{S.}~\bibnamefont{Jevrejeva}},
  \bibinfo{journal}{Nonl.\ Proc.\ Geophys.} \textbf{\bibinfo{volume}{11}},
  \bibinfo{pages}{561} (\bibinfo{year}{2004}).

\end{thebibliography}
\end{document}